\input harvmac.tex \input epsf 
\newcount\figno \figno=0
\def\fig#1#2#3{
\par\begingroup\parindent=0pt\leftskip=1cm\rightskip=1cm\parindent=0pt
\baselineskip=11pt \global\advance\figno by 1 \midinsert \epsfxsize=#3
\centerline{\epsfbox{#2}} \vskip 12pt {\bf Figure \the\figno:}  #1\par
\endinsert\endgroup\par } \def\figlabel#1{\xdef#1{\the\figno}}
\def\encadremath#1{\vbox{\hrule\hbox{\vrule\kern8pt\vbox{\kern8pt
\hbox{$\displaystyle #1$}\kern8pt} \kern8pt\vrule}\hrule}} \batchmode
\font\bbbfont=msbm10 \errorstopmode \newif\ifamsf\amsftrue \ifx\bbbfont\nullfont
\amsffalse \fi \ifamsf \def\IR{\hbox{\bbbfont R}} \def\IZ{\hbox{\bbbfont Z}}
\def\IF{\hbox{\bbbfont F}} \def\IP{\hbox{\bbbfont P}} \else \def\IR{\relax{\rm
I\kern-.18em R}} \def\IZ{\relax\ifmmode\hbox{Z\kern-.4em Z}\else{Z\kern-.4em
Z}\fi} \def\IF{\relax{\rm I\kern-.18em F}} \def\IP{\relax{\rm I\kern-.18em P}}
\fi

  \def\tilde{\widetilde}
   
\def\tr{{\rm Tr ~}} \font\zfont = cmss10 
\def\ZZ{\hbox{\zfont Z\kern-.4emZ}} 


 

\def\journal#1&#2(#3){\unskip, \sl #1\ \bf #2 \rm(19#3) }
\def\andjournal#1&#2(#3){\sl #1~\bf #2 \rm (19#3) }

\def\frac#1#2{{#1\over#2}}

\def\inbar{\,\vrule height1.5ex width.4pt depth0pt}
\def\IC{\relax\hbox{$\inbar\kern-.3em{\rm C}$}} \def\IR{\relax{\rm I\kern-.18em
R}} \def\IP{\relax{\rm I\kern-.18em P}}    

\def\slash#1{\mathord{\mathpalette\c@ncel{#1}}} \overfullrule=0pt

\def\RR{{\cal R}}   
   
\def\ZZ{{\cal Z}}   
\def\underrel#1\over#2{\mathrel{\mathop{\kern\z@#1}\limits_{#2}}}

 \catcode`\@=12


\def\tr{{\rm tr}}        


\def\nsp{{NS$^\prime$}}

\Title{\vbox{\rightline{hep-th/9706047} \rightline{IASSNS--HEP--97/61}}}
{\vbox{\centerline{Chiral Symmetry from Type IIA Branes}}}


\centerline{Amihay Hanany and Alberto Zaffaroni} \smallskip{\it
\centerline{School of Natural Sciences} \centerline{Institute for Advanced
Studies} \centerline{Princeton, NJ 08540, USA}} \centerline{\tt hanany@ias.edu
\tt zaffaron@ias.edu}



\noindent

We propose a mechanism in which, using an eightbrane, a sixbrane ends on a NS
brane in type IIA superstring theory.  We use this mechanism to construct $N=1$
supersymmetric gauge theories in four dimensions with chiral matter localized in
different points in space.  Anomaly cancellation for the gauge theories is
satisfied by requiring RR charge conservation for the various type IIA fields.
The construction allows us to study a curious phase transition in which the
number of flavors in supersymmetric QCD depends on the value of the ten
dimensional cosmological constant.  These phenomena are related to the fact that
every time a D8-brane crosses a NS brane, a D6-brane is created in between them.

\Date{5/97}

\lref\kut{S.  Elitzur, A.  Giveon, and D.  Kutasov, {\it Branes and $N=1$
Duality in String Theory}, hep-th/9702014.}


\lref\tatar{R. Tatar {Dualities in 4D Theories with Product Gauge Groups from
Brane Configurations}, hep-th/9704198.}

\lref\kuttt{S.  Elitzur, A.  Giveon, D.  Kutasov, E.  Rabinovici and A.
Schwimmer, {\it Brane Dynamics and N=1 Supersymmetric Gauge Theory},
hep-th/9704104.}

\lref\jon{N.  Evans, C.  V.  Johnson and A.  D.  Shapere, {\it Orientifolds,
Branes, and Duality of 4D Gauge Theories}, hep-th/9703210.}

\lref\hw{A.  Hanany and E.  Witten, {\it Type IIB Superstrings, BPS Monopoles,
And Three-Dimensional Gauge Dynamics}, IASSNS-HEP-96/121, hep-th/9611230.}

\lref\gp{E.  G.  Gimon and J.  Polchinski, {\it Consistency Conditions for
Orientifolds and D-Manifolds}, Phys.Rev.  D54 (1996) 1667, hep-th/9601038.}

\lref\dbl{M.  Berkooz, M.  R.  Douglas, R.  G.  Leigh, {\it Branes Intersecting
at Angles}, hep-th/9606139, Nucl.Phys.  B480 (1996) 265-278.}

\lref\kleb{ U.  H.  Danielsson, G.  Ferretti and I.  R.  Klebanov, {\it Creation
of Fundamental Strings by Crossing D-branes}, hep-th/9705084;

O.  Bergman, M. R.  Gaberdiel and G.  Lifschytz, {\it Branes, Orientifolds and
the Creation of Elementary Strings}, hep-th/9705130.}

\lref\barb{J.  L.  F.  Barbon, {\it
Rotated Branes and $N=1$ Duality}, CERN-TH/97-38, hep-th/9703051.}

\lref\witten{E.  Witten, {\it Solutions Of Four-Dimensional Field Theories Via M
 Theory}, hep-th/9703166.}

\lref\brodie{J.H.  Brodie and A.  Hanany, {\it Type IIA Superstrings, Chiral
Symmetry, and $N=1$ 4D Gauge Theory Duality}, hep-th/9704043}

\lref\dug{M.  R.  Douglas, {\it Branes within Branes}, hep-th/9512077}

\lref\ahup{O.  Aharony and A.  Hanany, unpublished.}

\lref\ah{O.~Aharony and A.~Hanany, {\it Branes, Superpotentials and
Superconformal Fixed Points}, hep-th/9704170.}

\lref\pol{J.  Polchinski, {\it Dirichlet-Branes and Ramond-Ramond Charges},
Phys.  Rev.  Lett.  75 (1995) 4724}

\lref\alwis{S.  P.  de Alwis, {\it Coupling of branes and normalization of
effective actions in string/M-theory}, hep-th/9705139}

\lref\witt{E.  Witten, {\it String Theory Dynamics In Various Dimensions},
Nucl.Phys.  B443 (1995) 85, hep-th/9503124.}

\lref\polstro{J.  Polchinski and A.  Strominger, {\it New Vacua for Type II
String Theory}, Phys.Lett.  B388 (1996) 736.}

\lref\karch{I.  Brunner and A.  Karch, {\it Branes and Six Dimensional Fixed
Points}, hep-th/9705022.}

\lref\barak{B.  Kol, {\it 5d Field Theories and M Theory}, hep-th/9705031.}

\lref\lll{K. Landsteiner, E. Lopez, D. A. Lowe, {\it N=2 Supersymmetric Gauge
Theories, Branes and Orientifolds}, hep-th/9705199.}

\lref\tel{A. Brandhuber, J. Sonnenschein, S. Theisen, S. Yankielowicz,
{\it M Theory And Seiberg-Witten Curves: Orthogonal and Symplectic Groups},
hep-th/9705232.}

\lref\oog{J. de Boer, K. Hori, H. Ooguri, Y. Oz, Z. Yin,
{\it Mirror Symmetry in Three-Dimensional Gauge Theories, SL(2,Z) and D-Brane
Moduli Spaces}, hep-th/9612131.}

\newsec{Introduction}

By now it is clear that D-branes \pol\ are a very useful tool in constructing
gauge theories in various dimensions and with various supersymmetries.  Such a
construction
has proved to be a very useful tool in studying the dynamics of the gauge
theories constructed.  In \hw\ it was proposed how to construct a brane
configuration which describes gauge theories with 8 supercharges in three
dimensions.  This construction can be T-dualized and give similar configurations
in various dimensions as was exploited in \refs{\witten, \ah, \karch, \barak}.
Application of the methods in \hw\ were performed also in \oog.
In addition see some recent generalization of the work of \witten\ in
\refs{\lll,\tel}.

\lref\costas{C. Bachas, M. R. Douglas, M. B. Green, {\it Anomalous Creation of
Branes}, hep-th/9705074.}

\lref\ted{A. Brandhuber, J. Sonnenschein, S. Theisen, S. Yankielowicz,
{\it Brane Configurations and 4D Field Theory Dualities}, hep-th/9704044.}

\lref\vafa{S. Katz, C. Vafa {\it Geometric Engineering of N=1 Quantum Field
Theories},  hep-th/9611090.}

\lref\vafadue{M. Bershadsky, A. Johansen, T. Pantev, V. Sadov, C. Vafa {\it
F-theory, Geometric Engineering and N=1 Dualities}, hep-th/9612052.}

\lref\vafatre{C. Vafa, B. Zwiebach, {\it N=1 Dualities of SO and USp Gauge
Theories and T-Duality of String Theory}, hep-th/9701015.}

\lref\vafaquattro{H. Ooguri, C. Vafa {\it Geometry of N=1 Dualities in Four
Dimensions} hep-th/9702180.}

\lref\ahn{C. Ahn, K. Oh {\it Geometry, D-Branes and N=1 Duality in Four
Dimensions I}, hep-th/9704061.}

\lref\ahndue{C. Ahn {\it Geometry, D-Branes and N=1 Duality in Four Dimensions
II}, hep-th/9705004.}

\lref\ahntre{C.Ahn, R. Tatar, {\it Geometry, D-branes and N=1 Duality in Four
Dimensions with Product Gauge Group}, hep-th/9705106.}

\lref\tatar{R. Tatar, {\it Dualities in 4D Theories with Product Gauge Groups
from Brane Configurations}, hep-th/9704198.}

\lref\bhoy{J. de Boer, K. Hori, Y. Oz, Z. Yin,
{\it Branes and Mirror Symmetry in N=2 Supersymmetric Gauge Theories in Three
Dimensions}, hep-th/9702154.}

The configuration in \hw\ was generalized to construct gauge theories with
4 supercharges in four dimensions in \kut.
Applications of this approach can be found in \refs{\bhoy, \jon, \brodie, \ted,
\kuttt, \tatar}.
A different approach to study the same class of theories, consisting in
wrapping D-branes on Calabi-Yau cycles, is analyzed in \refs{\vafa, \vafadue,
\vafatre, \vafaquattro, \ahn, \ahndue, \ahntre}.

The theories constructed so far with 4 supercharges are vector like.  It is very
natural to try and construct chiral gauge theories and study their dynamics
using the new brane methods. It should be mentioned that chiral matter from
branes was constructed in \dbl\ and the anomaly was canceled by space-time
fields. This will not be the approach we will take in this paper.

A first step toward understanding the construction of chiral gauge theories was
done in \brodie\ where the chiral symmetry of supersymmetric QCD was identified
in terms of the branes.  When sixbranes parallel to NS branes meet in space, the
sixbranes can break and form two different gauge groups.  In \brodie\ it was
also proposed how chiral multiplets arise when three branes meet at the same
point in space time.  Concretely, the proposal was that when a semi-infinite D6
brane meets a D4 brane on the world-volume of a NS fivebrane, there is a
massless chiral multiplet in the spectrum.  Such a brane system is far from
being described by the usual D-brane methods and thus, unfortunately it is not
possible to calculate directly how such a multiplet arises.  However we can look
for some consistency checks which will give support for this proposal.

In \ah\ a rather nontrivial check on the proposal in \brodie\ was performed.
The superpotentials of various brane configurations where calculated.  It was
found that precisely at the points where the superpotential is zero - where
chiral symmetry is expected - sixbranes can actually meet parallel NS branes and
then break.

The approach we will take in this paper is to assume that this proposal is
correct and study its consequences.  In doing this we find some more
consistency conditions which supports this proposal.
We find new interesting phenomena which arise due to this assumption.
This also provides the next step towards
constructing a wide class of theories with chiral matter.

A first question which arises when a semi-infinite sixbrane ends on a NS
fivebrane in type IIA superstring theory is that there is lack of gauge
invariance for the RR field which carries the D6-brane charge.  In other words,
the charge of the D6-brane can not flow anywhere when the D6-brane ends on a NS
brane.  We will first address this question and solve it by introducing
D8-branes.  Such branes do not break further the supersymmetry as will be
discussed in section 2.1 .  These D8-branes can be considered to be located far
away from the system of D4-branes to avoid new massless states in the four
dimensional theory.  Their effect is to create a cosmological constant for the
type IIA string theory.  We will show that for non-zero value of the
cosmological constant a semi-infinite sixbrane can end on a NS fivebrane.  In
this setup, if a D4-brane ends on the NS-brane and meets the D6-halfbrane, there
is a massless chiral multiplet in the spectrum.

However a single chiral multiplet is anomalous in four dimensions and thus we
expect some inconsistencies to arise.  Indeed this is what happens and whenever
we try to construct gauge theories with anomalous matter there is some RR-charge
violation in the string theory setup.  Conversely, only configurations with no
charge violations give rise to anomaly free theories in four dimensions.
Starting with configurations which obey the RR-charge conservation the theory
may undergo a phase transition which naturally preserve the RR-charge.  Thus the
theories beyond the phase transition remain anomaly free.  Chiral multiplets can
however be supported near one of the boundary of the D4-brane.  The KK reduced
four dimensional theory is anomaly free, but one may have local anomalies
supported on the boundaries of the segment on which the theory is defined.  All
anomalies, global or localized on the boundaries, must cancel in a consistent
theory, and we indeed show how the presence of a cosmological constant induces
Chern-Simons terms on the D4-worldvolume which precisely cancel the anomalies
on the boundaries.

The cancellation of the RR charge in spacetime, for the configurations
considered
in this paper, corresponds to a non-chiral four dimensional fields content.
If we move background branes around in space, the four dimensional theory may
undergo phase transitions.  We consider here only final configurations which are
still non-chiral.  The matter is however localized in different points in space
according to its chirality.  In one of these phase transitions, we connect $N=1$
gauge theories with different number of flavors.  This phenomenon is obtained by
changing a real parameter in spacetime (the position of a D8-brane) and has no
field theoretical counterpart in the four dimensional theory, where a complex
parameter is required to give mass to the quarks.  The number of flavors is
connected to the value of the ten dimensional cosmological constant, which is
indeed quantized.

The cancellation of the RR charge gives informations also about the
six-dimensional gauge theory living on the world-volume of the D6-branes.  We
show in section 3, that the charge conservation in spacetime exactly reproduces
the anomaly cancellation condition for the six-dimensional gauge theory.  In the
meanwhile, we point out that, whenever a D8-branes crosses a NS-brane, a
D6-brane is created in between them.  This phenomenon is T-dual to the one
originally discussed in \hw , where a D3-brane is created in between D5 and
NS-branes.  It is also U-dual to the creation of a fundamental string between D0
and D8-branes recently considered in \kleb\ and other examples of branes
creation in \costas .

The paper is organized in the following way.  In section 2, we first review the
proposal for chiral symmetry in \brodie\ and extend it to $SO$ and $USp$ gauge
groups.  We discuss then how a D6-brane can be finite without violating the RR
charge conservation.  The D6-brane must end on a D8-brane or on a NS-brane in
the presence of non-zero cosmological constant.  In section 3, the system
consisting of D6, D8 and NS-branes is applied to the study of six-dimensional
$N=1$ gauge theories.  The phenomenon for which a D6-brane is created every time
a D8-brane crosses a NS-brane is crucial for the correct understanding of the
six-dimensional theory.  In section 4, we apply the set-up of section 2 and 3 to
four-dimensional $N=1$ gauge theories, showing how to localize the chiral matter
in different points of the space.  A curious phase transition in which the
number of flavors depends on the value of the cosmological constant is also
presented.  Section 5 contains some brief concluding remarks.  The appendix
contains the normalizations for the RR charges and for the parameters in the
massive type IIA Lagrangian.

\newsec{Half D6-branes, massive type IIA and chirality}

\subsec{A dictionary for the branes configuration} \subseclab{\secsusy}

In this section we introduce the configurations and present the setup of the
construction.  We recall the dictionary of how states arise in various
configurations.  Next we recall the problem of how to identify the global chiral
symmetry in $N=1$ gauge theory models built with configuration of branes in type
IIA string theory.  The set-up is the same as in \kut, where it was introduced
to explain Seiberg's duality, adapting a construction first appeared in \hw .

The ingredients of the construction are Dirichlet (D) fourbranes, Dirichlet
sixbranes and Neveu-Schwarz (NS) fivebranes in type IIA string theory.  They
share the space-time directions $(x^0, x^1, x^2, x^3)$ where the
four-dimensional $N=1$ gauge theory is realized.  The D6-branes and the
NS-branes are infinite in the remaining directions and can be considered as
infinitely heavy background branes.  The D4-branes are stretched between D6 or
NS-branes in the $x^6$ direction.  Thus they are finite in the $x^6$ direction
and their world-volume theory is on an interval.  After Kaluza-Klein reduction,
the D4-branes theory is a four-dimensional gauge theory in which the
world-volume fields of the D6 or NS-branes appears as background fields
associated with global symmetries.

More explicitly, the ingredients are:  \item{(1)} NS fivebrane with world-volume
$(x^0, x^1, x^2, x^3, x^4, x^5)$, which lives at a point in the $(x^6, x^7, x^8,
x^9)$ directions.  The NS fivebrane preserves supercharges of the
form\foot{$Q_L$, $Q_R$ are the left and right moving supercharges of type IIA
string theory in ten dimensions.  They are (anti-) chiral:
$\epsilon_R=-\Gamma^0\cdots\Gamma^9\epsilon_R$,
$\epsilon_L=\Gamma^0\cdots\Gamma^9\epsilon_L$.}  $\epsilon_LQ_L+\epsilon_RQ_R$,
with \eqn\nsfive{ \eqalign{\epsilon_L=&\Gamma^0\cdots\Gamma^5\epsilon_L\cr
\epsilon_R=&\Gamma^0\cdots\Gamma^5\epsilon_R.\cr }}

\item{(2)} \nsp\ fivebrane with world-volume $(x^0, x^1, x^2, x^3, x^8, x^9)$,
which lives at a point in the $(x^4, x^5, x^6, x^7)$ directions, preserving the
supercharges

\eqn\nsprime{ \eqalign{\epsilon_L=&\Gamma^0\Gamma^1\Gamma^2\Gamma^3
\Gamma^8\Gamma^9\epsilon_L\cr \epsilon_R=&\Gamma^0\Gamma^1\Gamma^2\Gamma^3
\Gamma^8\Gamma^9\epsilon_R.\cr }}

\item{(3)} D4-brane with world-volume $(x^0, x^1, x^2, x^3, x^6)$, which lives
at a point in the  $(x^4, x^5, x^7, x^8, x^9)$ directions and preserves
supercharges satisfying \eqn\dfour{
\epsilon_L=\Gamma^0\Gamma^1\Gamma^2\Gamma^3\Gamma^6\epsilon_R.  }

\item{(4)} D6-brane with world-volume $(x^0, x^1, x^2, x^3, x^7, x^8, x^9)$,
which lives at a point in the $(x^4, x^5, x^6)$ directions.  The D6-brane
preserves supercharges satisfying \eqn\dsix{
\epsilon_L=\Gamma^0\Gamma^1\Gamma^2\Gamma^3\Gamma^7\Gamma^8 \Gamma^9\epsilon_R.
} We will need in the following also $D6^{\prime}$-branes spanning $(x^0, x^1,
x^2, x^3, x^4, x^5, x^7)$ and D8 brane with world volume $(x^0, x^1, x^2, x^3,
x^4, x^5, x^6, x^8, x^9)$.

Each brane by itself would break ${1\over 2}$ of the supersymmetry, but it is
easily checked that the relations \nsfive-\dsix\ are not independent.  It was
found in \hw\ for a T-dual configuration that combining the relations \nsfive\
and \dfour\ leads to the relation \dsix.  Thus in the presence of the D4-branes
and the NS-brane, the addition of the D6-brane does not break further the
supersymmetry and there are 8 unbroken supersymmetries.  Once the NS$'$ brane is
introduced, the supersymmetry is reduced further to 4 unbroken supercharges.
Now it is possible to combine all possible supersymmetry conditions and look for
new types of branes for which the presence will not break supersymmetry further
\ahup.  For example combining equation \nsprime\ and equation \dfour\ leads to a
supersymmetry condition which is obeyed in the presence of D6$'$-branes.  A
combination of equation \nsprime\ and equation \dsix\ leads to a supersymmetry
condition which is obeyed in the presence of a D8-brane.  It is easy to check
that combination of all other supersymmetry conditions does not lead to new
types of branes other than the ones mentioned above \ahup.  There is of-course a
possibility to rotate the NS branes and the D6 branes by an arbitrary angle in
the 45 - 89 directions as discussed in \ah\ and applied in \brodie.  We will not
use this in the present paper but it would be interesting to study the effects
of such rotations on the low energy field theory dynamics.  As a result of all
these considerations, a $N=1$ supersymmetry is realized on the D4 world-volume.

The finite D4-branes can end on either 5-branes or 6-branes.  In principle, the
world volume of the D4 brane contains a $N=4$ $U(1)$ multiplet, but some
massless states are projected out by boundary conditions.  These conditions are
different whether the D4-branes end on 5-branes or 6-branes.  Let us review,
\hw, what fields survive in the different cases.

1) For a 4-brane between two NS 5-branes, there is a $N=1$ vector multiplet and
chiral multiplet in the adjoint representation living on the world volume theory
of the 4-brane (the adjoint chiral multiplet corresponds to motion of the
4-brane in the $(x^4,x^5)$ direction).  For $n$ coincident D4-branes, we get a
$SU(n)$ gauge theory\foot{As in \witten, the $U(1)$ multiplet is frozen by
requiring finite energy configuration on the 5-branes.}.

2) For a D4-brane between an NS 5-brane and an \nsp\ 5-brane there is only a
vector multiplet.  In general, the two 5-branes can be at an arbitrary angle
$\theta$, \dbl, in the directions $(x^4,x^5,x^8,x^9)$ without breaking any
further supersymmetry.  See also a related discussion in \barb .  The mass of
the adjoint chiral multiplet that appeared in item one is proportional, for
small angles, to the angle between the two 5-branes.  For $n$ such D4-branes we
get $SU(n)$ gauge theory.

3) For a D4-brane between two D6-branes, only scalars fields survive.  These
 scalars are associated with motion of the D4-brane along the $(x^7,x^8,x^9)$
 coordinates together with the $A_6$ component of the gauge field.  The $A_6$
 component is compact with a radius proportional to the gauge coupling.  These
 four scalars are conveniently written in terms of two chiral multiplets
 $x_8+ix_9$; $x_7+iA_6$.

4) For a D4-brane between a D6-brane and an NS 5-brane there are no massless
moduli which contribute to the low energy field theory.  In general, this is
true for any NS5-brane and D6-brane that are not parallel.  For more than one
D4-branes between a D6-brane and a NS brane - an S-configuration - the system is
conjectured to break supersymmetry \hw.

5) For a 4-brane between a D6-brane and \nsp\ 5-brane there is one chiral
multiplet $x_8+ix_9$, which corresponds to motion of the D4 brane in between the
two branes.  In general there will be one chiral multiplet which is associated
with the motion of a D4-brane between parallel D6-brane and a NS brane.

In addition to the above massless states, there are other fields which come from
open strings stretching between the 4-branes and the 6-branes.  These strings
give fields in the fundamental representation of the gauge group, $Q$ and
$\tilde Q$, the quarks.  They carry also indices in the fundamental of the gauge
group living on the world-volume of the 6-branes.  Since the 6-branes are
infinite in the $(x^7,x^8,x^9)$ directions, this gauge group is frozen from the
four-dimensional point of view and appears as a global symmetry.

\fig{A system of semi-infinite D6 branes which end on a NS$'$ brane and finite
D4 branes which end on the NS$'$.  A chiral multiplet is conjectured to arise
from strings stretched between the two D branes.}  {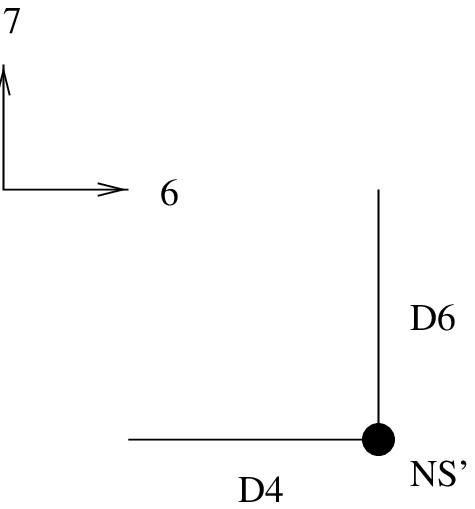}{6 truecm}
\figlabel\chiral

Following \brodie, we can now add one more case for which there are massless
chiral moduli. For completeness of the dictionary we state this case here
however one should bare in mind that the possible charge violation and anomaly
should be avoided as discussed in detail in the rest of this section.

6) When a D4 brane ends on a NS$'$ brane which is a boundary of
a semi-infinite sixbrane there is a massless chiral multiplet in four
dimensions.  This multiplet arises due to strings stretched between the D6 and
the D4 branes.  Only one orientation of the strings contributes to the massless
states which leads to only one chiral multiplet, the other orientation being
projected out.  In the usual case in which the D6 branes are infinite both
orientations of the strings contribute massless states and thus we have two
chiral multiplets.  When $N_c$ branes end to the left (in the $x^6$ direction)
of the NS$'$ brane and $N_f$ D6 branes end on top (in the $x^7$ direction), as
in figure \chiral, the chiral multiplet transforms in the $(N_c,\bar N_f)$
representation.  Other cases, like when the $N_c$ branes end to the right,
transform in an obvious way keeping in mind that the orientation of the strings
which contribute to the massless states remains the same.

\subsec{Chiral symmetry}

We now turn to some applications of the dictionary above.  The simplest example,
SQCD, can be easily constructed by stretching $N_c$ D4-branes between a NS
5-brane and a \nsp\ 5-brane, with $N_f$ D6-branes in between the two NS-branes.
The distance in the $(x^4, x^5)$ plane between the D6 and the D4 branes is the
complex mass parameter for the quarks $Q,\tilde Q$.  When all the masses are
zero, the D6-branes meet the D4 branes.  Tuning further the $x^6$ positions of
the D6-branes they can coincide in spacetime and form an enhanced gauge
symmetry.
The $SU(N_f)$ gauge theory on the D6 world-volume becomes a $SU(N_f)$ global
symmetry rotating the quarks.  A transition to the Higgs branch is now possible
in field theory.  In the brane language, it corresponds to splitting the
D4-branes between the D6-branes which are now at the same point.  A consistent
picture of the resulting moduli space is obtained if one takes in consideration
the s-rule, discovered in \hw :

{\it A configuration in which a NS-brane and a D6-brane are connected by more
than one D4-brane is not supersymmetric}.

SQCD has actually an $SU(N_f)_L\times SU(N_f)_R$ chiral global symmetry, whose
diagonal $SU(N_f)$ we identified with the D6 world-volume gauge theory.  The
remaining part of the chiral symmetry is apparently lacking in the brane
construction.  This is related to the fact that most pieces in this construction
are {\it locally} N=2, at least for what regards the field content.  The quarks
Q and $\tilde Q$, for example, are generated by the local interaction between a
D4 and D6 brane system which has $N=2$ supersymmetry in four dimension and
(regarding the field content) hardly knows about the existence of distant NS
branes that further break the supersymmetry to N=1.  In the $N=2$ supersymmetric
QCD case, there is indeed a term in the superpotential, \eqn\wneqtwo{W = \lambda
QX\tilde Q.}  where X is the adjoint chiral superfield partner of the gauge
fields under $N=2$ supersymmetry.  $\lambda$ is fixed to $\sqrt 2$ by
supersymmetry.  This term implies that when we give a vacuum expectation value
to the adjoint field, the hypermultiplets receive a mass.  This term also breaks
the global chiral symmetry from $SU(N_f)_R \times SU(N_f)_L$ to $SU(N_f)$.
Turning off the coefficient $\lambda$ in \wneqtwo, breaks the $N=2$
supersymmetry to $N=1$ while restoring the chiral symmetry.

A mechanism for identifying the missing chiral symmetry was proposed in \brodie:
the chiral symmetry is restored when the 6-branes touch a parallel 5-brane.  To
make, say a \nsp\ brane and a D6 brane coincide, we need to tune the three
coordinates $(x^4,x^5,x^6)$.  In \brodie\ it was proposed that a massless
multiplet comes down in the world volume $(x^0,x^1,x^2,x^3,x^8,x^9)$ which is
mutual to both branes.  The supersymmetry in the \nsp -D6 system is $(0,1)$ in
six dimensions, and the only supermultiplet which can become massless by tuning
only three parameters is a vector multiplet.  The distance in the
$(x^4,x^5,x^6)$ directions gets the interpretation of a FI parameter for this
$U(1)$ gauge field.

The only direction that the 6-branes and the 5-brane do not share is $x^7$.
Therefore, a 6-brane can have a boundary on the 5-brane in the $x^7$ direction.
Thus the proposal is that when $N_f$ 6-branes touch a parallel 5-brane, they
split in half, and the $(x^0,x^1,x^2,x^3,x^8,x^9)$ world volume theory becomes
$SU(N_f)_R\times SU(N_f)_L$.  Strings stretching from the $N_c$ 4-branes to the
$N_f$ 6-branes to the right in $x^7$, are the quarks $Q$ charged under $(\bf
{N_c,N_f,1})$.  Strings stretching from the $N_c$ 4-branes to the $N_f$ 6-branes
to the left in $x^7$, are the quarks $\tilde Q$ charged under $(\bf {\bar
N_c,1,N_f})$.  Moving the 6-branes off the 5-brane in the $(x^4,x^5)$ direction
breaks the six dimensional symmetry spontaneously while breaking the four
dimensional chiral symmetry explicitly, as it should since motion in the
$(x^4,x^5)$ corresponds in the field theory to giving the quarks a mass.
Breaking the 4-branes on the 6-branes is only possible if the 6-branes move off
the 5-brane in the $(x^8,x^9)$ directions.  This corresponds to Higgsing, and in
agreement with field theory, Higgsing breaks chiral symmetry.  Actually, from
the point of view of the six brane observer, the motion of the D6 branes in the
456 directions correspond to moving on a Higgs branch while the point when the
D6 branes touch the NS$'$ brane is the origin of this Higgs branch.  There is no
Coulomb branch for vector multiplets in six dimensions and thus there is no
other motion from this point, associated with breaking.

The chiral symmetry proposal was further checked and explored in \ah.  There the
setup was mostly in the context of three dimensional $N=2$ gauge theories
realized in terms of branes which is a T-dual version of the configurations we
discuss here.  In \ah\ superpotentials were calculated for a class of brane
configurations.  It was found that in cases where the superpotential vanishes,
that is when chiral symmetry is expected, the transition of the type described
in \brodie\ and above are possible.  The calculation in \ah\ thus serves as a
non-trivial consistency check for the proposal of \brodie.  One of the purposes
of this paper is to provide further consistency checks for this proposal.

\subsec{SO(n) and Sp(n) gauge groups} \subseclab{\othergroup}

In this section we want to discuss the changes in the previous construction when
applied to $SO(2N_c)$ and $USp(2N_c)$ gauge groups with $N_f$ flavors.  The
field theory analysis predicts a maximal global symmetry $SU(2N_f)$.  Only a
subgroup is visible in the standard construction with infinite D6-branes.  We
show that, using semi-infinite D6-branes, the maximal global symmetry becomes
manifest. \foot{A.H. would like to thank discussions on related issues with
Per Kraus and Jaemo Park.}

$SO(n)$ and $Sp(n)$ gauge theories can be realized by introducing suitable
orientifold planes \jon\kuttt .  We will consider in this section the simplest
case in which the orientifold plane is parallel to the D4-branes and we will
call it O4.  It does not break any further supersymmetry but it creates images
for all the branes in the directions $x^4,x^5,x^7,x^8,x^9$ transverse to the
D4-plane.  The space-time operation of reflecting the coordinates
$x^4,x^5,x^7,x^8,x^9$ is accompanied by the world-sheet parity $\Omega$.  If we
project out the Chan-Paton factors of $2n$ branes ($n$ physical branes plus $n$
images) with a symmetric unitary matrix we obtain an $SO(2n)$ gauge group for
the D4-branes, while if we project with an antisymmetric unitary matrix we
obtain the $USp(2n)$ gauge group\foot{In a superstring compactification, only
the the symmetric projection is allowed.  In fact, the antisymmetric projection
corresponds to the unusual case in which branes and orientifold planes charges
have the same sign and, therefore, when the space transverse to the branes is
compact there is no way to cancel the charge or, equivalently, the tadpoles.  In
our case, in which the transverse space is not compact, both projections are
allowed.}.  The symmetric case allows to consider also an odd number $2n+1$ of
D4-branes, realizing an $SO(2n+1)$ gauge theory.  In this case, one of the
D4-branes has no image and it is stucked at the orientifold point.

The D6-branes, which meet orthogonally the D4-branes, are projected by a matrix
which is antisymmetric if the one which projects the D4-branes is symmetric, and
vice versa \gp .  In this way, the global symmetry manifest in the brane
construction for $SO(2N_c)$ gauge theories is $USp(2N_f)$ and for $USp(2N_c)$
gauge theories is $SO(2N_f)$.  This result is consistent with the field
theoretical expectations for an $N=2$ gauge theory.  When the gauge theory is
$N=1$, from the field theory point of view, we expect a maximal $SU(2N_f)$
global symmetry.  The $N=1$ case is realized using two different orientations of
solitonic
five-branes, the NS and the \nsp . As discussed in subsection \chiral, we expect
to see the maximal global symmetry when the D6-branes touch the \nsp\ and split.

We will consider only the case of an $SO(2N_c)$ gauge theory, for simplicity.
The $SO(2N_c)$ gauge group is unbroken if all the $2N_c$ D4-branes are at the
orientifold point, which we take at $x^4,x^5,x^7,x^8,x^9=0$.  All the
hypermultiplets are massless when the D6-branes touch the D4-branes along the
orientifold plane.  In a generic point in the $x^6$ direction, the global
symmetry is $USp(2N_f)$.  In general, the Chan-Paton factor for $2N_f$ branes is
a generic hermitian matrix in $U(2N_f)$.  The $\Omega$ projection reduces it to
a matrix in $USp(2N_f)$.

 However, when the D6-branes are on top of the \nsp -brane, they can split,
giving a total of $4N_f$ half D6-branes.  The orientifold projection acts now in
a totally different way.  The world-volume field living on a D6-halfbrane which
fills the positive $x^7$ direction is identified with the world-volume field on
a D6-halfbrane which fills the negative $x^7$ direction, but no projection is
needed on the $2N_f\times 2N_f$ hermitian matrix which describes the Chan-Paton
factors of the half D6-branes.  In this way, we find a manifest $SU(2N_f)$
global symmetry, as predicted by the field theoretical analysis.

\subsec{RR Charge Conservation}

It would be natural now to conjecture that a configuration in which only half of
the D6-brane exists, say the semi-infinite D6-brane to the right in $x^7$, would
correspond to a chiral model, with only the quark $Q$.  Unfortunately, this
configuration violates the RR 7-form charge conservation.  The RR charge flux of
a Dp-brane ending on a (NS or RR) q-brane is usually absorbed by some fields on
the q-brane world-volume.  This works only when $q > p$.  In our case there is
no way in which the NS 5-brane can absorb the flux of the D6-brane.

\lref\green{M.  Green, J.  A.  Harvey and G.  Moore, {\it I-Brane Inflow and
Anomalous Couplings on D-Branes}, Class.  Quant.  Grav.  14 (1997) 47,
hep-th/9605146.}

This violation of charge conservation is actually a good sign from the point of
view of the four dimensional field theory.  A gauge theory with only one charged
field $Q$ is anomalous and thus inconsistent.  This means that a violation of
the RR 7-form charge is translated to an anomaly for the four dimensional gauge
theory.  Thus when looking for anomaly free theories we will look for string
theory brane configurations which do not violate charge conservation as they
must in order to be consistent\foot{There exist also systems of branes which
realize anomalous field theories on the common intersection \green .
The anomaly is cancelled by an in-flow of charge from the background branes.
We will not consider this situation in this paper.}. 

\subsec{D8-branes and massive type IIA supergravity} \subseclab{\massive}
In this section we present a mechanism for compensating the RR 7-form charge
violation induced by a semi-infinite D6-brane.  Let us start by writing
explicitly this charge violation.

The D6-brane is charged under the type IIA 7-form $A^{(7)}$, the magnetic dual
of the type IIA $U(1)$ form $A^{(1)}$.  The Bianchi identity for $A^{(1)}$ in
the presence of a semi-infinite D6-brane, say, in the positive $x^7$ direction,
reads:  \eqn\bianchi{dF^{(2)}=d*F^{(8)}= \theta (x^7)\delta^{(456)}} Taking the
differential of this equation, we derive the inconsistent relation
$0=\delta^{(4567)}$ which reveals that the RR charge is not conserved in such a
configuration.

The simplest way to compensate the RR charge violation is to make the D6-brane
end on a D8-brane.  As we saw in \S\secsusy\ the supersymmetry still allows the
existence of a D8-brane which is a point in the $x^7$ direction.  The RR charge
flux is then absorbed by the $U(1)$ gauge field which lives on the D8
world-volume and which sees the end of the D6-brane as a magnetically charged
5-brane which we will call a magnetic monopole.

The existence of a D8-brane at $x^7=0$ would modify the right hand side of
\bianchi\ in the following way \eqn\bianchiuno{dF^{(2)}=d*F^{(8)}= \theta
(x^7)\delta^{(456)} - F^{(2)}_{D8}\delta (x^7)} where $F^{(2)}_{D8}$ is the
$U(1)$ gauge field strength which lives on the D8 world-volume.  Differentiating
\bianchiuno\ we now get the statement that the D6-brane endpoint is a
magnetically charged monopole for the D8 gauge field:
$dF^{(2)}_{D8}=\delta^{(456)}$.

The modification to \bianchi\ follows from the couplings to the RR background
fields existing on the world-volume of a Dp-brane \dug, \eqn\RR{\int dx^{p+1}
C\wedge \tr e^{(F-B)}} where C is the formal sum of all the RR background forms,
$F$ is the world-volume gauge field, $B$ is the background NS-NS antisymmetric
tensor and it is to be understood that the world-volume integral selects the
$(p+1)$-form in the expansion.  In the case of a D8-brane the relevant coupling
which affects the equation of motion for $A^{(7)}$ (the Bianchi identity for
$A^{(1)}$) is $A^{(7)}\wedge F^{(2)}_{D8}$.

As discussed in section \secsusy , four dimensional chiral matter arises when
the D6-halfbrane boundary coincide with the \nsp -brane on which the D4-brane
ends.  If a D8-brane is placed at the same point in $x^7$ to make the half
D6-brane
consistent, there is a new multiplet in the D4-worldvolume associated with the
4-8 strings.  To avoid the creation of new multiplets in the four-dimensional
spectrum, it would be useful to have a different mechanism for reabsorbing the
RR flux of the D6-brane.

The only other way to modify \bianchi\ without involving new extra branes would
be to get a contribution to this equation directly from the bulk type IIA
supergravity Lagrangian.  An easy check reveals that there are no terms in type
IIA supergravity which can modify \bianchi\ in a way which compensates the RR
charge violation.  At this point it would seem hopeless to construct chiral
theories as described in the last section.  However a term which can compensate
the charge conservation exists in the massive type IIA \lref\rom{L.  Romans,
Massive N=2A supergravity in ten-dimensions, Phys.  Lett.  B169 (1986) 374.}
supergravity \rom\ in the form of the coupling \eqn\massi{-m\int dx^{10} B\wedge
*F,} where $m$ is the mass parameter for the type IIA, related to the
ten-dimensional cosmological constant.  In a massive type IIA background,
\bianchi\ now would read:  \eqn\bianchidue{dF^{(2)}=d*F^{(8)}= \theta
(x^7)\delta^{(456)} - mH.}

Differentiating this equation, we get $mdH=\delta^{(4567)}$.  We conclude that
the RR charge flux of a half D6-brane can be reabsorbed in the massive type IIA
background by a NS 5-brane coinciding with the boundary of the half D6-brane.
This is exactly the configuration of branes which was conjectured in \brodie\ to
make manifest the chiral global symmetry of SQCD, as discussed at the end of the
previous section.

Let us analyze this proposal further, showing how the massive type IIA arises
naturally in the brane \lref\wp{J.  Polchinski and E.  Witten, {\it Evidence for
heterotic - type I string duality}, Nucl.  Phys.  {\bf B460} (1996) 525.}
\lref\ps{J.  Polchinski and A.  Strominger, {\it New vacua for type IIA string
theory}, Phys.Lett.  B388 (1996) 736, hep-th/9510227.}  {\it phenomenology}
\wp\ps .  In this discussion, we will discover also that the two approaches for
compensating the RR charge of a half D6-brane (with a D8-brane or with the
massive type IIA background) are not so unrelated as they may seem at a first
look.  The D8-brane is charged under the RR form $A^{(9)}$.  The $A^{(9)}$ gauge
potential is exceptional in type IIA string, since it has no dual potential.
Its field strength does have a dual field strength but the field equations
restrict this to be a constant, $m$.  This constant is essentially the square
root of the cosmological constant appearing in the massive IIA supergravity
theory \rom .  The equation of motion for $A^{(9)}$ in the type IIA string in
the presence of D8-branes (which we take to be a point, say, in the $x^7$
direction) implies that $F^{(10)}$ is piece-wise constant function away from the
D8-branes with a jump of one unit when we cross one D8-brane \wp.  This results
in a step-like non-zero cosmological constant, proportional to the square of the
constant value of $*F^{(10)}$ .  In the region of space between two different
D8-branes the bulk fields are conveniently described by the massive type IIA
supergravity.  In the units that we use in this paper, the parameter $m$ is an
integer which jumps by $n$ units when one meets $n$ coincident
D8-branes\foot{The appropriate normalizations for fields and charges are
discussed in the appendix, where the exact relation between $m$ and the number
of D8-branes and their charge is derived.  In this paper, all the RR charges are
set to 1, unless explicitly said.}.

The D8-branes, whose introduction we tried to avoid, are apparently coming back
in the game, since the only known supersymmetric backgrounds that solve the IIA
field equations for $m\ne 0$ require the presence of eight-branes.  However, we
may assume that the D8-branes are positioned far away at infinity in the
direction $x^7$ and that their only role is to provide a non-zero cosmological
constant.

In this set-up, half D6-branes can exist provided that they end on parallel
NS-branes and the deficit of RR charge is compensated by a suitable value of the
cosmological constant.  Let us restate the condition which follows from
\bianchidue\ reintroducing the charge $\mu$ for the D6 and the NS branes, for
the configuration in which $n_6$ D6-branes end on $n_{ns}$ NS-branes,
\eqn\charge{n_6\mu_6 = mn_{ns}\mu_{ns}.}  The parameter $m$ is equal to the
$\mu_8$ charge of the D8-branes times the number of such D8-branes.  It is an
highly non-trivial check that the actual value of the various charges allows a
solution in terms of three integers, the number of D6, NS and D8 branes.  The
condition becomes, \eqn\chargeuno{n_6=mn_{ns}} where $m$ is now normalized, up
to an integer factor, to correspond to the number of D8-branes present at
infinity.  The check of value of the charges will be performed in the appendix.

As we saw, the two approaches for reabsorbing the RR flux of an half D6-brane
are
closely related.  We can always create a cosmological constant simply by moving
a decoupled D8-brane sitting at infinity to the other side of our system.
According to the previous discussion, a semi-infinite D6-brane can now end on
the \nsp\ brane.  To be more precise, a finite D6-brane {\it must} be now
stretched between the \nsp -brane and the D8-brane.  Every time a D8-brane
crosses a \nsp -brane, a D6-brane is created between them.  This is nothing but
a ``T-dual'' version of the effect discovered in \hw.  This will be the subject
of section 3.  To simplify the discussion we will consider the system consisting
only of D6, D8 and \nsp -branes, which realizes a six dimensional gauge theory.

\newsec{More about D6, D8 and \nsp -branes} \subsec{Gauge Theory in
six-dimensions} \subseclab{\sixdim}

Before discussing the application of this mechanism to the understanding of the
chiral symmetry in four dimensional SQCD realized as a configuration of
D4-branes, we want to apply this construction to a simpler model describing
six-dimensional gauge theories.  We will obtain a consistency check for our
approach by deriving the anomaly restrictions for an $SU(N)$ six-dimensional
gauge theory, and, as a by-product, we will determine the actual relation
between the parameter $m$ and the number of D8-branes.

Thus, we learn that in order to understand some aspects of chiral gauge theories
in four dimensions we need to study the dynamics of six dimensional theories.  A
``T-dual'' version of this approach was considered in \ah.  There, the study of
five dimensional gauge theories which can be realized by branes in type IIB was
performed.  It was found that each such model gives rise to a three dimensional
gauge theory by stretching a D3 brane between a pair of fivebrane polymers.  We
will use the same strategy here by studying first six dimensional theories which
can be realized in terms of branes and then use such systems to the study of
four dimensional theories using D4-branes.

Consider a system with \nsp, D6 and D8-branes.  Their world-volume is as
indicated in the dictionary in section \secsusy .  In the directions $(x^0, x^1,
x^2, x^3, x^8, x^9)$ which are common to all the branes a $N=1$ six-dimensional
gauge theory is realized.  In the set-up of section \secsusy, this
six-dimensional gauge theory must be reinterpreted as the background global
symmetry of the gauge theory living on the D4-branes.  For the moment, we will
forget about D4-branes and we will consider the six-dimensional theory on its
own.  This configuration with \nsp, D6 and D8-branes can be considered as the
T-dual of the system originally proposed in \hw\ and has been studied recently
in \karch.

\fig{A system of D6 NS$'$ and D8 branes. The D6 branes are stretched along the
$x^7$ direction and are represented by horizontal lines, the NS$'$ branes are
localized in this direction and are
represented by points and the D8 branes are localized in this direction and are
represented by vertical lines.}  {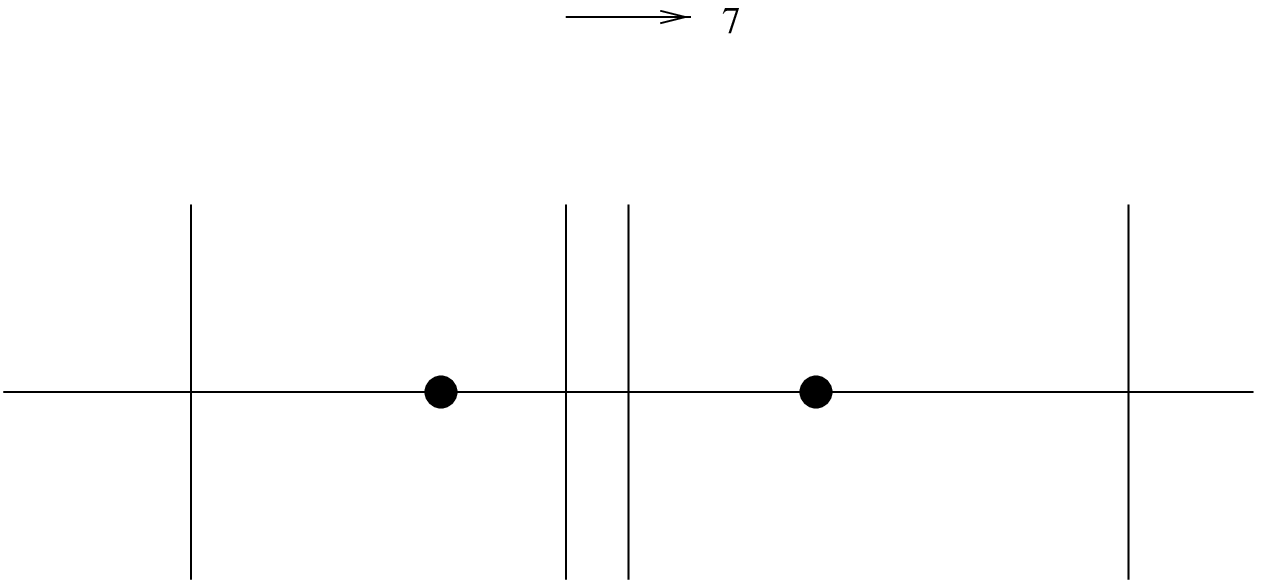}
{10 truecm} \figlabel\simsixd

We can picture, as in figure \simsixd, a straight line, the $x^7$ direction,
along
which the D6-branes are stretched.  The \nsp -branes are points on the line on
which the D6-branes are assumed to end.  The D8-branes intersect the line and
the 68-open strings give rise to hypermultiplets on the world-volume of the
D6-branes.

Consider the simplest example with only two \nsp -branes \karch\ and N D6-branes
along the $x^7$ direction.  We can now break the D6-branes between the \nsp ,
considering the theory on the semi-infinite D6-branes on the left or on the
right of the \nsp -branes as a frozen theory with zero coupling constant.  From
the field theory point of view, we are dealing with an $SU(N)$ gauge theory on
the
world-volume of the N D6-branes stretched between the two \nsp -branes, with 2N
hypermultiplets coming from the open strings which connect the D6-branes with
the semi-infinite D6-branes on the left or on the right.

\lref\seib{N.  Seiberg, hep-th/9609161; {\it Phys.  Lett.}  {\bf B390} (1996)
753.}  \lref\dani{U.  Danielsson, G Ferretti, J.  Kalkkinen and P.  Stjernberg,
hep-th/9703098.}

The matter content of a $N=1$ six-dimensional gauge theory is highly constrained
by the gauge anomaly \seib\dani .  If the gauge group has an independent fourth
order Casimir element (and this is the case for SU(N), at least for $N>3$), the
gauge anomaly cancellation constrains the matter content.  For SU(N), the number
of flavors must be twice the number of colors.  As a general rule, it is
believed that the charge conservation for the bulk fields corresponds to the
anomaly cancellation on the world volume of the branes.  In our example, the
conservation of the RR charge for the D6-branes corresponds to the statement
that
the number of flavors is $2N$, which is exactly the condition for the anomaly
cancellation in field theory.  It should be noted that even for number of
flavors equal to twice the number of colors, the anomaly must be cancelled by
the introduction of a tensor field \seib .  Here, a tensor field in the
six-dimensional theory is automatically provided by each of the \nsp -branes
\karch .  One of the two tensors parametrizes the center of mass motion of the
system and it is decoupled from the theory.  The scalar $\phi$, partner of the
dynamical tensor under $N=1$ supersymmetry, is associated with the distance
between the two \nsp -branes.  The same distance is also the gauge coupling of
the theory.  The fact that the two quantities coincide corresponds to the
well-known fact \seib\ that in the Lagrangian \eqn\kar{\frac{1}{g^2} F^2_{\mu
\nu} + (\partial \Phi)^2 + \sqrt{c} \Phi F_{\mu \nu}^2} ${1\over g^2}$ can be
absorbed in $\phi$.

\fig{Six dimensional theory realized in terms of NS$'$, D6 and D8 branes.  There
are $N$ D6 branes stretched in between two NS$'$ branes.  $n_l (n_r)$ D6 branes
end
to the left (right) of the NS$'$ branes.  In addition there are $n$ D8 branes in
between the two NS$'$ branes.  The configuration describes $SU(N)$ gauge theory
with $n+n_l+n_r$ flavors coupled to one tensor multiplet.}  {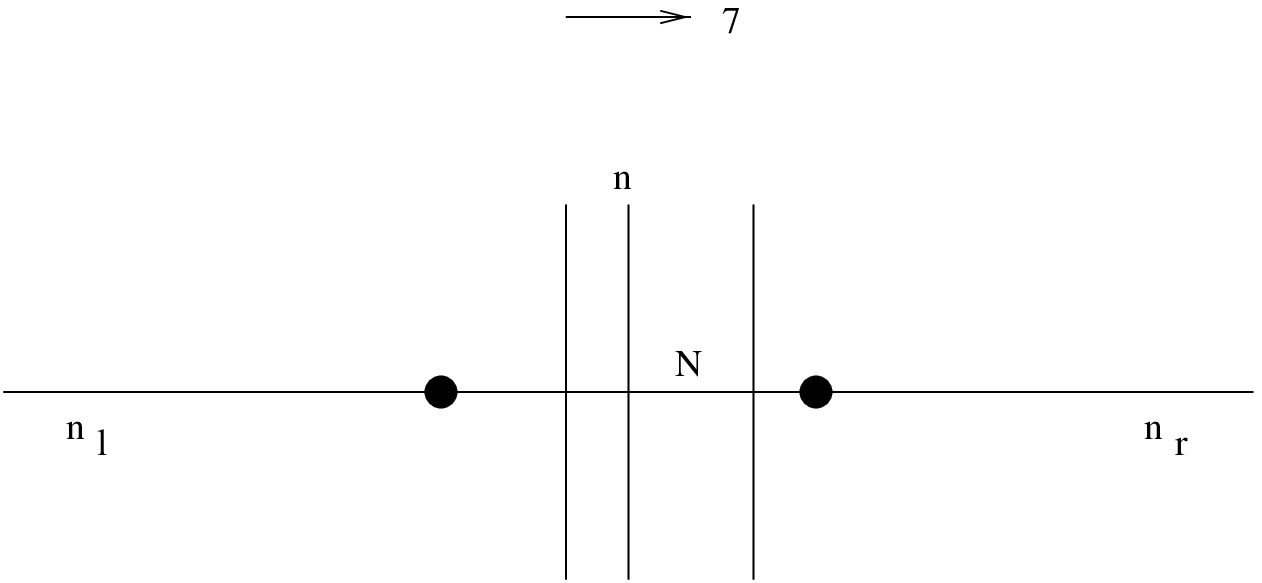}{10
truecm} \figlabel\sixd

We can easily generalize this construction by introducing, as in figure \sixd,
$n$ D8-branes which intersect the $x^7$ line in points between the two \nsp
-branes and considering now a different number $n_l$ ($n_r$) of semi-infinite
D6-branes on the left (right) of the \nsp -branes.  The mismatch of charge
between the left and the right of each \nsp -brane can be compensated by the
cosmological constant induced by the D8-branes.  We will assume that there are
other D8-branes at infinity in the $x^7$ direction in such a way that the value
of the parameter $m$ on the left of the first \nsp -brane is $p$.  Every time we
meet a D8-brane on the $x^7$ axis the value of $m$ jumps by 1 unit.  This means,
for example, that the value of $m$ on the right of the second \nsp -brane is
$p+n$.

We expect that the charge conservation still gives the anomaly cancellation
constraint.  The two charge conservation equations now
read:\eqn\cha{\eqalign{N-n_l+p=&0\cr n_r-N+p+n=&0}} and we indeed get the
relation that the total number of flavors $N_f=n_l+n_r+n$ must be equal to $2N$.
Note that the enhanced non-Abelian flavor symmetry in this case is
$SU(n_l)\times SU(n)\times SU(n_r)$ for the case in which all three families of
branes meet.  This is a subgroup of the expected flavor symmetry which is
$SU(N_f)$.  The full $SU(N_f)$ symmetry is visible, for example, if all flavors
are given by D8 branes in between the two NS$'$ branes.  That is for $n_l=n_r=0$
and $n=2N$ when the $x^7$ positions of the D8 branes coincide we see a $SU(N_f)$
symmetry.

\subsec{Creation of a sixbrane} \subseclab{\sixbrane}

The exact value of the $x^7$ position of the D8-branes did not enter in the
previous discussion.  This is obviously related to the fact that hypermultiplets
can not be massive in six-dimensions, so we do not expect parameter associated
with them.  The role of their position is analogous to the $x^6$ position of the
D5-branes in \hw ; this parameter is not visible as a deformation of the quantum
field theory, but it may be associated to non-trivial phase transition
corresponding to moving a D8-brane on the other side of a \nsp -brane.  Exactly
as in \hw, where a D3-brane appears when a D5-brane passes through a NS brane,
we expect that D6-branes can be created when a D8-brane passes through a \nsp
-brane.  The charge conservation tells us whether this happens or not.

Before studying some six dimensional properties related to the motion of the D8
branes let us mention a fact which will be useful later.  We can always think
that the semi-infinite D6-branes end on D8-branes located at infinity.  This can
be done without loss of generality due to the following process.  Suppose we
have an infinite D6 brane and let us put a D8 brane in some very large value of
$x^7$.  The two branes touch and thus the D6 brane can break on the D8 brane.
One side is stretched to the origin of $x^7$ and the other is stretched far at
large $x^7$ values.
Now we can move this latter semi-infinite D6 brane to large 456 values.  Thus we
are left with a D6 brane which ends on a D8 brane at very far values of $x^7$.
The gauge theory which is located around the origin of $x^7$ is not
affected from this process.

Let us consider the
simplest example, $n_l=n_r=N$ and no D8-branes $n=0$ (a configuration consistent
with a zero cosmological constant).  This is a $U(N)$ gauge theory with $2N$
hypermultiplets.  We use what we have said in the last paragraph and assume that
the semi-infinite D6-branes end on D8-branes located at infinity.  Now move one
of these D8-branes located at $x_7=-\infty$ to the right of the first \nsp
-brane.  One semi-infinite D6-brane disappears and one D8-brane is now in
between the two \nsp -branes.  The theory is still a $U(N)$ gauge theory with
$2N$ hypermultiplets.  No new D6-brane is appeared. The relations
$n_l=N-1, n=1, n_r=N$ are consistent since the apparent lack of charge at the
position of the
first \nsp -brane is compensated by the value $m=-1$ created by the D8-brane on
the region of space on its left.  But when the D8-brane moves
to the right of the second \nsp-brane a D6-brane is created between the \nsp\
and the D8-brane.  $n_l=N-1, n=0, n_r=N+1$ is indeed the only configuration
which satisfies the charge conservation. This follows from the
fact that  the cosmological
constant has been turned on ($m=-1$) in the whole region which contains the two
\nsp -branes. We can now pull the D8-brane to the
positive infinity and we remain with the same gauge theory, $U(N)$ with 2N
hypermultiplets, but with a different arrangement of semi-infinite branes on the
left and on the right, $n_l=N-1,
n_r=N+1$.  

Since the configuration of NS, D8 and D6-branes is T-dual to the system
considered in \hw , we expect that the creation of D6-branes is also regulated
by a linking number argument.  We want now to show that the RR charge
conservation and the linking number argument are actually equivalent.

Let us briefly review the linking number argument.  The presence of a \nsp\
brane
generates a non-trivial background for the $B$ field.  This affects the D8
world-volume theory since the D8 $U(1)$ gauge field mixes with the bulk field
$B$
in such a way that $F_{D8}-B$ is the gauge invariant curvature.  We start with a
general configuration of D6-branes ending on D8-branes.  The endpoints
of the D6-branes look like monopoles with centers in $(x^4_i,x^5_i,x^6_i)$ for
the D8 world-volume gauge field, \eqn\link{d(F_{D8}-B) = dF_{D8}-H =
\sum_i\delta_{D6}(x^4_i,x^5_i,x^6_i)} If we integrate this equation over the
$R^3$ spanned by $(x^4,x^5,x^6)$, $\int dF_{D8}$ gives the total magnetic charge
for the $U(1)$ gauge field.  This quantity is a constant and because it can be
measured at infinity it is unaffected by the motion of the branes in
spacetime.  We conclude that the sum of $\int_{D8}dx^4dx^5dx^6 H$ on the
world-volume of the D8-brane with the total net number of D6-branes ending on
the D8 must be conserved.  $\int_{D8}dx^4dx^5dx^6H$ still depends on the
position in $x^7$ of the D8-brane and it is easily checked that it has a
different sign whether the D8-brane is on the left or on the right of the \nsp\
brane.  The jump
of 1 unit\foot{The actual value of the jump was derived in \hw\ using the
geometrical interpretation of the constant magnetic chrage as the linking number
invariant when the D5 and NS-worldvolume is conveniently compactified. The value
follows also from the equations of motion when normalizations for fields and
charges are carefully taken into account.  This will be done explicitly in the
appendix.  As already noted, there is a non-trivial cancellation of RR charge
quanta which results in an integer.}  which occurs when the D8-brane passes
through the \nsp -brane is compensated by the creation of a D6-brane.  This is the
phenomenon that we explained before using the RR charge conservation for the
type IIA 1-form.  This is not a mere coincidence occurring in this specific
example and the two arguments in six dimensions are equivalent.  This means that
we do not have to double the number of constraints which regulates the allowed
configurations and the phase transitions, but simply use the one that we prefer.
The equivalence can be explicitly checked for a generic configuration of D6, D8
and \nsp -branes, once one remembers that, for the S-rule, no more than one
D6-brane can end on a D8-brane if the other end lives on a \nsp -brane.
It is easy to see that   the linking
number and the charge conservation argument coincide, provided $m(x^7)$ is
assumed to jump
by 1 unit at the location of each D8-brane.  This assumption was used constantly
in this section and will be proved in the appendix.  The previous discussion can
also be considered as a further check of this assumption.

\newsec{Chiral symmetry in four dimensions} \subsec{Localized chiral matter}
\subseclab{\chiral}

We now return to SQCD in four-dimension.  We will show that using the mechanism
described in the previous sections, we can make the chiral symmetry manifest on
the brane configuration by localizing matter of different chirality in different
points.

\fig{Localization of chiral matter in $N=1$ supersymmetric QCD.  In figure (a)
the usual brane configuration which describes this theory is sketched.  Vertical
lines represent D6 branes which end on D8 branes that are located very far in
the $x^7$ direction.  Hypermultiplets arise at points were sixbranes and
fourbranes meet.  Figure (b) describes the resulting configuration after the D8
branes move to the other side of the $x^7$ coordinate.  Chiral multiplets are
localized at the two ends of the fourbranes, on the fivebranes world volume.  }
{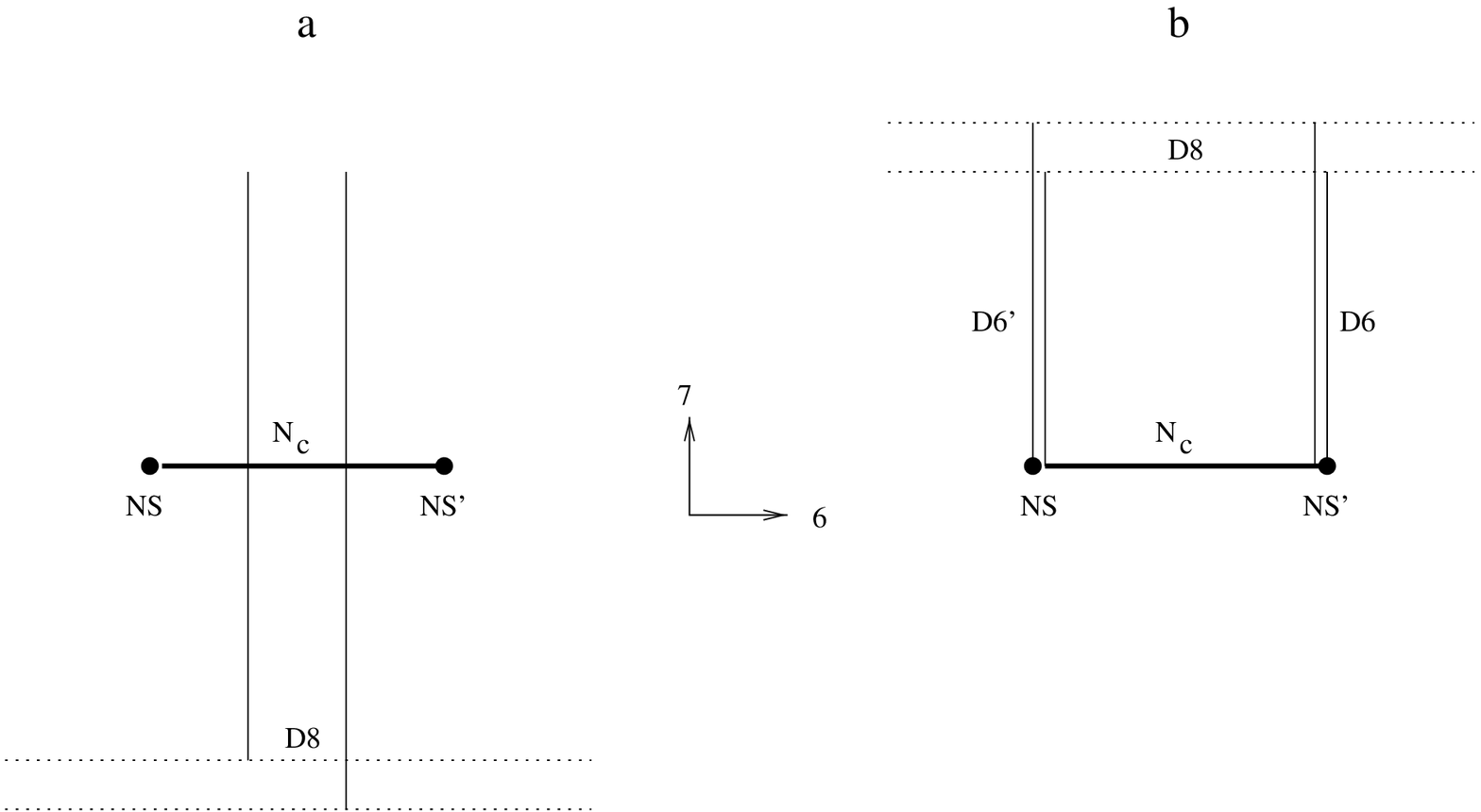}{12 truecm} \figlabel\SQCD

$SU(N_c)$ SQCD is realized by stretching $N_c$ D4-branes in the $x^6$ direction
between a NS and a \nsp -brane.  When the $N_f$ D6-branes are tuned to touch the
\nsp -brane, each of them splits into two semi-infinite D6-branes.  Each of the
D6-halfbranes supports a $SU(N_f)$ gauge group and we can see the full
$SU(N_f)_l\times SU(N_f)_r$ chiral symmetry of the four dimensional gauge
theory.  We want to give further evidence for this proposal \brodie\ by using
the construction of the previous sections.

We can imagine, without loss of generality, that the D6-halfbranes end on
D8-branes located at infinity in the $x^7$ direction.  This can be done by the
following process which has the interpretation of a Higgs mechanism for the six
dimensional theory.  We can take $N_f$ D8 branes and break each of the D6
branes.  Now there are two half six branes per each D8 brane.  One half is
connected to a NS$'$ brane far away and by the rules of section \secsusy\ and
\hw\ can not move.  The other half is free to move along the 456 directions and
can be taken far away in these directions.  Note that we need $N_f$ D8 branes in
order to avoid S-configurations which are not supersymmetric \hw.  The resulting
configuration is depicted in figure \SQCD (a).  The cosmological constant can be
tuned (by adding other far away D8 branes) to be zero in the region of spacetime
which contains our brane system.  Now move the D8-branes which are located at
negative infinity in $x^7$ to positive infinity in $x^7$.  They have to pass
through the NS and \nsp -branes and thus D6-branes can be created according to
the discussion in the previous section.  The only difference is that if a
Dirichlet 6-brane is created with an end on the NS-brane, it must be parallel to
it, so it must be what we called a $D6^\prime$ -brane.  The value of the
cosmological constant is now $m=N_f$.  It is easy to enforce the charge
conservation.  In the anomaly free final configuration, shown in figure \SQCD
(b), there are $N_f$ D6-halfbranes ending on the \nsp -brane and $N_f$
D6$'$-halfbranes ending on the NS-brane.  Both of them are along the positive
$x^7$ axis.  The apparent lack of RR charge at both the NS and \nsp\ location is
compensated by the non-zero value of the cosmological constant.

The gauge theory on the world-volume of the D4-brane is still $SU(N_c)$ with
$N_f$ fields $Q$ in the fundamental and $N_f$ fields $\tilde Q$ in the
anti-fundamental, but now $Q$ is supported near the NS end of the D4-brane,
while $\tilde Q$ is supported near the \nsp\ end.  The global symmetry is
manifestly $SU(N_f)_l\times SU(N_f)_r$, and the two chiral factors are now
supported by D6-halfbranes at different positions.  This gives evidence to the
proposal in \brodie\ for manifestly realizing the chiral symmetry by splitting
the D6-branes.

Charge conservation in spacetime should correspond to anomaly cancellation on
the branes.  The final gauge theory is manifestly anomaly free, but the anomaly
cancellation is more subtle and it is worthwhile to comment on it.  The
world-volume quantum field theory on the D4-brane is actually a five-dimensional
gauge theory defined on $R^4$ times a segment.  A five-dimensional theory can
not have anomaly.  However the two boundaries are four-dimensional and the
boundary values of the spinors can give anomalies.  In the KK reduction all the
massive fields do not contribute to the anomaly, and the global anomaly is given
by the massless fields.  The theory for these massless fields is $SU(N_c)$ with
$N_f$ fields $Q$ in the fundamental and $N_f$ fields $\tilde Q$ in the
anti-fundamental for both the cases $m=0$ and $m\ne0$ and it is anomaly free.
However the anomaly must cancel also locally on each boundary.  In the $m\ne0$
case each boundary supports only the fields $Q$ or the fields $\tilde Q$, so the
anomaly is apparently not cancelled.  \lref\town{M.  B.  Green, C.  M.  Hull, P.
K.  Townsend, {\it D-brane Wess--Zumino actions, T-duality and the cosmological
constant} Phys.  Lett.  B382 (1996) 65, hep-th/9604119} However, as noted in
\town , in the presence of a cosmological constant there are extra terms on the
world-volume of a D-brane.  In particular, on a D4-brane there is a Chern-Simons
term \eqn\green{m\int dx^5 \omega_5} where $\omega_5$ is the Chern-Simons 5-form
with the property that $d\omega_5=\tr F^3$.  Under a gauge transformation,
$\delta\omega_5=d(F\wedge F)$.  We see that such a term is not gauge invariant
when the theory is defined on a segment, and gives an anomaly supported on the
boundary, \eqn\boun{\delta S_{D4} = m\int dx^5 d(F\wedge F) = m\int dx^4(F\wedge
F)|_1 - m\int dx^4(F\wedge F)|_2} where 1 and 2 refers to the two boundaries.
We see that this anomaly has exactly the sign and the magnitude necessary to
cancel the anomaly of the fields $Q$ and $\tilde Q$ supported on the two
boundaries\foot{The anomaly in \boun\ clearly cancels in the KK analysis and
does not contribute to the global anomaly.}.

This contribution from the Chern-Simons terms is always present when the
cosmological constant is non-zero.  For what regards anomalies, it is harmless 
for infinite branes.  But when the D-brane is defined on a segment, it produces
anomalies localized at the boundaries.  Starting with a D4-brane stretched
between NS-branes in the presence of a cosmological constant and nothing else,
we would find an anomalous theory and we would start looking for chiral
matter localized at the boundaries, discovering the necessity of half D6-branes.
This anomaly argument provides further support for the proposal \brodie\ of how
chiral matter arises.

The final configuration with non-zero cosmological constant indicates clearly
the origin of the chiral symmetry on the D6 halfbranes.  The fields of different
chirality are now supported on different halfbranes and it is tempting to
speculate that with this mechanism we could construct chiral models and even
realize a smooth process that connect non-chiral and chiral gauge theories in a
Seiberg's duality.  Example of this kind do actually exist in the literature.

\subsec{A curious phase transition} \subseclab{\phase} In the SQCD example of
the previous section, the original and final quantum field theories were
actually the same.  It is easy to produce an example in which, by moving a
D8-brane, we can connect two theories with different field content.

\fig{A phase transition in which the number of flavors for a gauge theory are
affected by the ten dimensional cosmological constant.  In figure (a) there are
$N_c$ fourbranes between a NS brane and a NS$'$ brane.  A D8 brane is in the
large negative $x^7$ position.  In figure (b) the D8 brane moves to the far
positive $x^7$ position and creates two D6 branes which connect to the two NS
branes.  A D6 is created from passing the NS$'$ brane and a D6$'$ is created
when passing the NS brane.}  {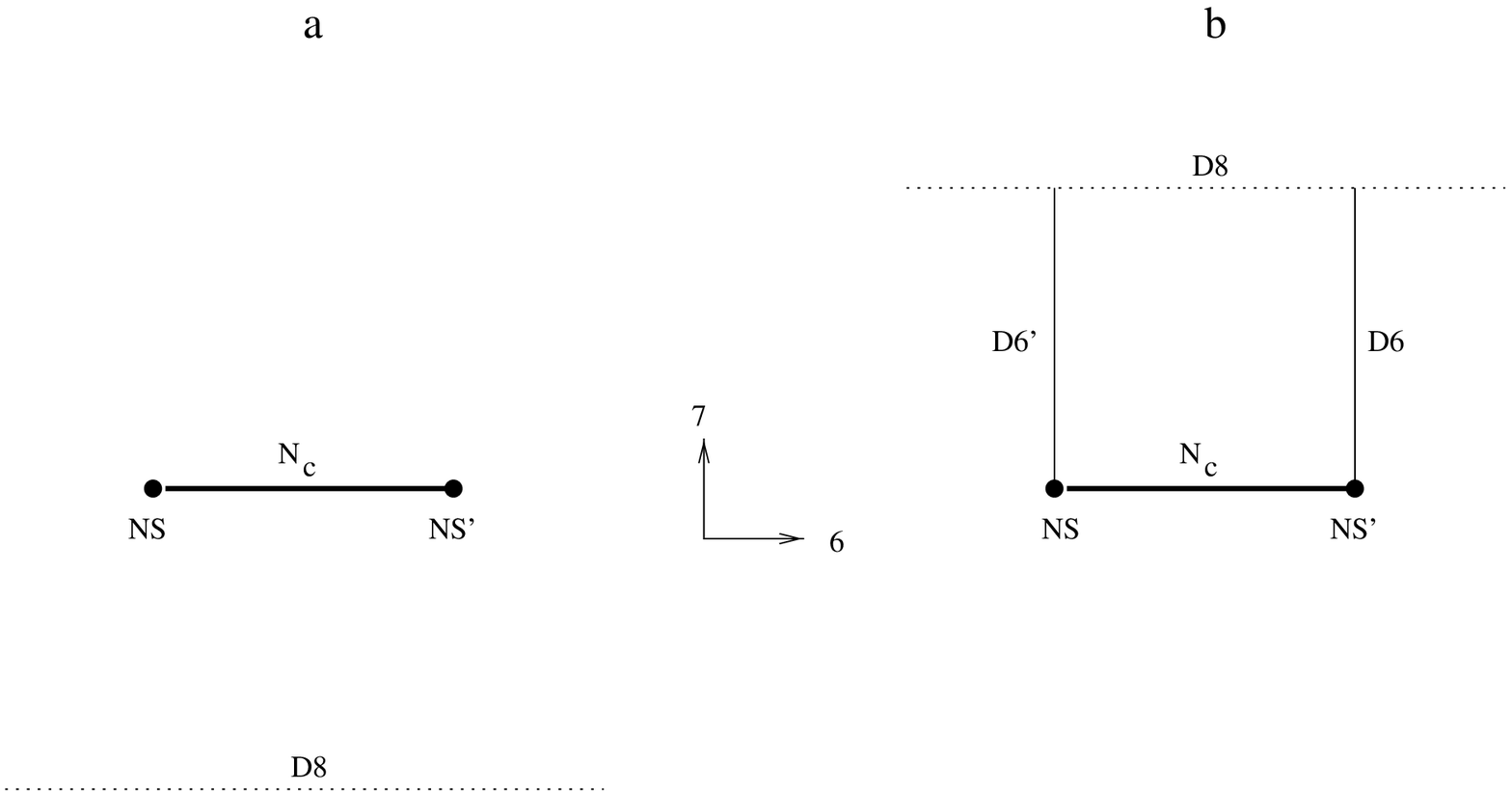}{12 truecm} \figlabel\addflavor

Consider $N_c$ D4 branes connecting a NS branes to a \nsp -brane.  Let the two
NS branes be positioned at the origin of the $x^7$ direction.  This
configuration describes $N=1$ supersymmetric $SU(N_c)$ gauge theory with no
matter.  Consider, as in figure \addflavor (a), a D8 brane positioned at a very
far negative $x^7=m_0$ position.  Such a D8 brane causes a jump in the value of
the cosmological constant.  \eqn\cosm{\eqalign{m(x^7)&= -1\qquad x^7<m_0\cr
m(x^7)&= 0\qquad x^7>m_0.\cr}} The value of the cosmological constant is zero at
the origin in $x^7$ and so the four dimensional system is not affected by the
presence of the D8 brane far away.  Now move one D8-brane, as in figure
\addflavor (b), from the large negative position to a positive large position in
the $x^7$ direction.  The cosmological constant at the origin of the $x^7$
direction, where the NS branes are located, is now $m=-1$.  Our discussion from
section \sixbrane\ implies that a D6 and $D6^\prime$ -branes are created, ending
on the \nsp -brane and the NS-brane, respectively.  The apparent paradox that
there is no charge conservation on the NS branes where the D6 branes end, is now
avoided by the fact that the cosmological constant has changed is value, as
explained in section \massive.

We now use our proposed dictionary for matter content as in section \secsusy\
and find \brodie\ that the matter content is now of a field $Q$ in the
fundamental representation and a field $\tilde Q$ in the anti-fundamental
representation of the gauge group.  The field $Q$ comes from strings stretching
between the D6 brane and the D4 branes while the field $\tilde Q$ comes from
strings stretching between the D6$'$ brane and the D4 branes.  The theory is now
SQCD with one quark flavor.

We saw that we can change the number of flavors of the SQCD theory simply by
moving D8-branes on the other side of the system.  This corresponds to a change
in the value of the cosmological constant in the region where the D4-branes
live. The number of flavor depends on the cosmological constant, which is indeed
quantized.  The curious thing is that we have changed the number of flavors by
changing a real parameter in spacetime - the position of the D8-brane.  This
phenomenon has no known explanation in terms of the field theory living on the
D4-brane, since the only way to make the quarks massive in a $N=1$ theory
involves complex parameters.

\newsec{Conclusions}

\lref\poul{P. Pouliot, {\it Chiral Duals of Non-Chiral SUSY Gauge Theories},
Phys.Lett. B359 (1995) 108, hep-th/9507018;

P. Pouliot, M. J. Strassler, {\it A Chiral $SU(N)$ Gauge Theory and its
Non-Chiral $Spin(8)$ Dual} Phys. Lett. B370 (1996) 76, hep-th/9510228;

M. Berkooz, {\it The Dual of Supersymmetric SU(2k) with an Antisymmetric Tensor
and Composite Dualities}, Nucl. Phys. B452 (1995) 513, hep-th/9505067;

K. Intriligator, R. G. Leigh, M. J. Strassler, {\it New Examples of Duality in
Chiral and Non-Chiral Supersymmetric Gauge Theories},
Nucl.Phys. B456 (1995) 567, hep-th/9506148. }

We have shown how to localize chiral matter in different spacetime points in a
system of D-branes in the type IIA string which gives a realization of $N=1$
gauge theories in four dimensions.

The ultimate goal would be to realize chiral models and, eventually, study
Seiberg's dualities associated.  There exists in the literature examples of
chiral
models dual to non-chiral gauge theories \poul.  It would not be surprising that
the phase transitions considered in this paper can be generalized to cover these
cases, too.  Charge conservation simply says that if we start with an anomaly
free theory, we must end after some phase transition with another anomaly free
theory.  But it does not exclude the option to end up with a chiral theory, if
we started with a non-chiral one.

The main problem to solve for realizing chiral models is how to construct
theories
with chiral two-indices tensors.  In this paper, we have shown how to get rid of
chiral fields in the fundamental of the gauge group.  $SU(N)$ theories with
chiral fundamental are anomalous (and indeed we did not get them by phase
transition in this paper) unless we compensate the anomaly with appropriate
chiral two-indices tensors.  These tensors do not arise naturally in the brane
set-up of this paper.  Some new ingredient is required.  This issue is currently
under investigation.

\appendix{A}{Normalization of RR charges}

In this appendix we want to discuss the normalization for the RR and NS brane
charges appearing in \charge\ and the relation between $m$ and the number of
D8-branes.

We will use the following normalizations for $H$ and $F^{(2)}$ in the massive
type IIA Lagrangian,

\eqn\lagran{S_{IIA} = -{1\over (2\pi )^7\alpha^{\prime 4}}\int d^{10}x \left
(e^{-2\phi}{1\over 2}H\wedge *H\right ) + {1\over 2}(F^{(2)}-mB)\wedge
*(F^{(2)}-mB).}

With this normalization for $H$, the coupling to the fundamental string is
$(2\pi \alpha^\prime )^{-1}\int B$.

The coupling between Dp-branes and RR fields is as follows, \eqn\RR{S = {1\over
2}\int d^{10}x F^{(p+2)}\wedge *F^{(p+2)} +i\mu_p\int d^{p+1}A^{(p+1)}.}


The value of the RR charge was found in \pol\ and reads, \eqn\polc{\mu^2_p =
2\pi (4\pi^2\alpha^\prime )^{3-p}.}  They satisfy the Dirac quantization
condition, \eqn\dirac{\mu_p\mu_{6-p} = 2\pi .}

It was suggested in \polstro\ that the cosmological constant in the massive type
IIA theory is quantized.  This is consistent with the fact that the only known
supersymmetric solutions of the equations of motion require the existence of
D8-branes and that the parameter $m$ is proportional to the number of D8-branes.
A simple argument \polstro\ allows to determine the exact proportionality
coefficient.  Consider the field equation \eqn\equaz{{1\over (2\pi
)^7\alpha^{\prime 4}}d*(e^{-2\phi}H) = m*(F^{(2)} - mB)} and integrate it over
an 8-sphere surrounding a D0-brane.  The inconsistent result that the flux $\int
*F^{(2)}$ is zero when $m\neq 0$ is modified if we suppose that a fundamental
string ends on the D0-brane and intersects the 8-sphere (this is exactly what a
D-brane is:  a manifold on which the open strings end\foot{More precisely, it
was pointed out in \kleb\ that whenever there is a cosmological constant a
fundamental string {\it must} be attached to the D0-brane.  In general, when a
D8-brane crosses a D0-brane (creating a non-zero cosmological constant in the
region in which the D0-brane lives) a fundamental string is created in between
them.  This phenomenon is U-dual to the creation of a D6-brane in between D8 and
\nsp -branes considered in this paper.}  ).  Such a string contributes to the
equation of motion of $B$ with a delta-function times $(2\pi \alpha^\prime
)^{-1}$.  The condition is now $m\int *(F^{(2)}-mB) - (2\pi \alpha^\prime )^{-1}
=0$ or \eqn\quant{m\mu_0 = {1\over 2\pi \alpha^\prime }} Using formula \polc ,
we learn that $m=\mu_8$.  This means that the presence of one D8-brane changes
by 1 unit the value of $m$, a fact that we used constantly in the paper.

The last thing we need to show is that in formula \charge\ the quanta of RR and
NS charge exactly cancel, giving the relation between the number of D6, NS and
D8 branes in formula \chargeuno .  With the conventions of this appendix, the
relevant equation is \eqn\rel{n_8\mu_8dH = n_6\mu_6\delta^{(4)}} Using the
quantization condition \quant\ and the relation \dirac , this is equivalent to
\eqn\reldue{ {n_8dH\over 2\pi\alpha^\prime } = n_6\mu_o\mu_6\delta^{(4)} = 2\pi
n_6 \delta^{(4)}.}

The coupling of a NS-brane can be explicitly derived by a dimensional reduction
of the M-theory five-brane \alwis .  Consider the relevant part of the eleven
dimensional supergravity Lagrangian,

\eqn\eleven{ -{1\over (2\pi )^8\alpha^{\prime {9\over 2}}}\int \left (\sqrt{-G}R
+ {1\over 2} F_M^{(4)}\wedge * F_M^{(4)}\right ).}

The KK reduction on a circle gives the type IIA Lagrangian if we use the ansatz
\witt , \eqn\ants{ds^2=e^{-{2\over 3}\phi}G_{\mu\nu}dx^\mu dx^\nu + e^{{4\over
3}\phi} (dy-A^{(1)}_\mu dx^\mu )^2,\qquad A^{(3)}_M = A^{(3)} + B^{(2)}\wedge
dy.}  where $y$ is the eleventh dimension coordinate with periodicity
$2\pi\sqrt{\alpha^\prime}$.  The normalization for H in the reduced Lagrangian
is as in \lagran, while the RR fields need to be rescaled by a factor $(2\pi
)^{7/2}\alpha^{\prime 2}$.

The five-brane charge in M-theory \eqn\five{\mu_{5M}\int A_M^{(6)}} can be
easily determined by identifying the M five-brane wrapped around the eleventh
dimension with the D4-brane of type IIA\foot{Remember that the RR fields must be
rescaled to agree with our notations.}

\eqn\tension{2\pi\sqrt{\alpha^\prime}(2\pi )^{7/2}\alpha^{\prime 2}\mu_{5M}=
\mu_4}

The reduced (not wrapped) M five-brane must be identified instead with the type
IIA NS-brane.  Reducing the coupling \five\ we get\foot{Note that the definition
of the Poincare dual involves the metric.  The ten dimensional metric is
obtained from the eleven dimensional one by a Weyl rescaling.  This produces the
extra dependence on the dilaton.}  \eqn\nsbr{{e^{-2\phi}\over (2\pi
)^5\alpha^{\prime 3}}\int *B^{(6)}.}

The equation for the NS-brane is now, \eqn\nsbrane{dH = n_{ns}(2\pi
)^2\alpha^\prime } When substituted in \reldue , it gives the desired relation
$n_8n_{ns}=n_6$.  The quanta of R and NS charge have disappeared.  As noted in
the text, this is an highly non-trivial result.

\vskip 0.5in

\centerline{\bf Acknowledgments}

We would like to thank Ofer Aharony, John Brodie, Kentaro Hori and Gilad Lifschytz
for useful discussions.
The research of A.H. is supported in part by NSF grant PHY-9513835.
The research of A.Z. is supported in part by DOE grant DE-FG02-90ER40542 and by
the Monell Foundation.

\vskip 0.5in

\listrefs \end